\documentclass[12pt]{iopart}
\usepackage{iopams}  
\usepackage{bm}
\usepackage[dvips,dvipdfm]{graphicx}
\newcommand{\V}[1]{\bm{#1} } 
 
\newcommand{\Wh}[1]{\widehat {#1}} 

\newcommand{\Extr}[1]{ \mathop{\rm Extr}_{ #1 } }

\newcommand{\mR}{\mathbb{R}}
\newcommand{\mN}{\mathbb{N}}

\begin{document}

\title
{Weight space structure and analysis using a finite replica number 
in the Ising perceptron}

\author{Tomoyuki Obuchi\dag\footnote[3]{obuchi@stat.phys.titech.ac.jp} and 
Yoshiyuki Kabashima\ddag
}

\address{
\dag
Department of Physics, Tokyo Institute of Technology,\\  
Tokyo 152-8551, Japan \\
\ddag
Department of Computational Intelligence and Systems Science, \\
Tokyo Institute of Technology, Yokohama 226-8502, Japan\\
}

\begin{abstract}
The weight space of the Ising perceptron in which a set of random patterns is stored is examined using the generating function of the partition function 
$\phi(n)=(1/N)\log [Z^n]$ as the dimension of the weight vector $N$ tends to infinity, where $Z$ is the partition function and $\left [ \cdots \right ]$ represents the configurational average. We utilize $\phi(n)$ for two purposes, depending on the value of the ratio $\alpha=M/N$, where $M$ is the number of random patterns. For $\alpha < \alpha_{\rm s}=0.833 \ldots$, we employ $\phi(n)$, in conjunction with Parisi's one-step replica symmetry breaking scheme in the limit of $n \to 0$, to evaluate the complexity that characterizes the number of disjoint clusters of weights that are compatible with a given set of random patterns, which indicates that, in typical cases, the weight space is equally dominated by a single large cluster of exponentially many weights and exponentially many small clusters of a single weight. For $\alpha > \alpha_{\rm s}$, on the other hand, $\phi(n)$ is used to assess the rate function of a small probability that a given set of random patterns is atypically separable by the Ising perceptrons. We show that the analyticity of the rate function changes at $\alpha = \alpha_{\rm GD}=1.245 \ldots $, which implies that the dominant configuration of the atypically separable patterns exhibits a phase transition at this critical ratio. Extensive numerical experiments are conducted to support the theoretical predictions. 
\end{abstract}



\maketitle
\section{Introduction}\label{sec:intro}

The generating function (density) with respect to 
the partition function $Z$: 
\begin{eqnarray}
\phi(n)=\frac{1}{N} \log [Z^{n}] \  \ (n \in \mR), 
\label{moment}
\end{eqnarray}
plays a key role in research on disordered systems, 
where $N$ denotes the size of the objective system and 
$[\cdots]$ denotes the average over the 
quenched randomness. 
Assessing $\phi(n)$ for $\forall{n} \in \mR$ exactly is, 
in general, difficult, whereas the analytical evaluation 
for $n =1,2,\ldots \in \mN$, in conjunction with the use of 
the saddle point method as $N \to \infty$, is possible for a class of systems. 
This indicates that (\ref{moment}) can be practically evaluated by 
analytically continuing the expressions of $\phi(n) $
evaluated for $n \in \mN$ to $n \in \mR$, 
which is often referred to as the {\em replica method}. 

In most models of statistical mechanics of disordered systems, the
probability that free energy density, $-(1/N)\log Z$, 
will take a certain value $f$, $P(f)$, can be expressed 
in large deviation statistics as
\begin{eqnarray}
P(f) \sim \exp \left \{ N R(f) \right \}, 
\label{rate_function}
\end{eqnarray}
where $R(f) \le 0$ is often referred to as the {\em rate function}.
One of recent progresses of the replica theory is
the formation of a link between $R(f)$ and $\phi(n)$ \cite{Pari1,Naka,Obuc}. 
When $R(f)$ is a convex upward function, it can be assessed
from $\phi(n)$ being parameterized by $n \in \mR$ as 
\begin{eqnarray}
f(n)=-\frac{\partial \phi(n)}{\partial n}, \quad 
R(f(n))= \phi(n)-n\frac{\partial \phi(n)}{\partial n}. 
\label{phi_R}
\end{eqnarray}
This indicates that the {\em typical} value of $f$, which is characterized
by the condition $R(f)=(1/N) \log P(f)=0$, can be evaluated as
\begin{eqnarray}
f^*=-\lim_{n \to 0} \frac{\partial \phi(n)}{\partial n}, 
\label{standard_replica}
\end{eqnarray}
which is sometimes referred to as 
a replica trick formula. 
Equation (\ref{phi_R}) indicates that 
$n=1,2,\ldots$ corresponds to {\em atypical} samples of 
$R(f) < 0$ representing a small probability. 
This means that the replica trick can be regarded as a formula that infers
the behavior of typical samples by extrapolating the behavior for atypical samples. 

Another recent advance in the replica theory is 
the association between the complex structure of 
phase space and a formalism of one-step replica symmetry breaking (1RSB)
\cite{Boff,Mona1,Mona3}. In a number of systems that are subject to disordered interactions, 
the phase space is considered to be divisible into 
exponentially many disjoint sets as $N \to \infty$. 
Each of the disjoint sets is sometimes referred to as a pure state. 
Let us assume that 
the number of pure states specified by 
the free energy (density) value $f$, ${\cal N}(f)$, 
is scaled as
\begin{eqnarray}
{\cal N}(f) \sim \exp \left \{ N \Sigma(f) \right \}, 
\label{complexity}
\end{eqnarray}
where the exponent $\Sigma(f) \ge 0$ is referred to as the {\em complexity}. 
Saddle point evaluation of $\sum_{\gamma } 
\exp \left \{-N x f_\gamma \right \}$, 
where $x \in \mR$ is a certain control parameter and 
$\gamma$ and $f_\gamma$ are indices of a pure state and its free energy, 
respectively, 
indicates that $\Sigma(f)$ can be evaluated using 
another generating function $g(x)=(1/N)\log \left (
\sum_{\gamma } 
\exp \left \{-N x f_\gamma \right \} \right )$ as
\begin{eqnarray}
f(x)=-\frac{\partial g(x)}{\partial x}, \quad 
\Sigma(f(x))=g(x)-x\frac{\partial g(x)}{\partial x}, 
\label{1RSB_complexity}
\end{eqnarray}
which is parameterized by $x$ as long as $\Sigma(f)$ is convex upward. 
This formalism is defined for each sample of quenched randomness and, 
therefore, has nothing to do with $\phi(n)$. 
However, recent studies have revealed that typical $g(x)$ (more precisely,
$(1/x)g(x)$) over the 
quenched randomness can also be assessed from $\phi(n)$ by evaluation of 
(\ref{standard_replica}) under Parisi's 1RSB ansatz,
handling the 1RSB parameter $x$ as a control parameter. 

The concepts of the two exponents $R(f)$ and $\Sigma(f)$ are
different in that $R(f) \le 0$ represents a small probability of
atypical samples, whereas $\Sigma(f) \ge 0$ represents a large number of 
pure states that occur for typical samples. However, 
the formal similarity of (\ref{phi_R}) and (\ref{1RSB_complexity})
indicates that there might be a relationship between these two exponents. 
In fact, when the 1RSB solution of $\phi(n)$, 
$\phi_{\rm 1RSB}(n)$, is assessed using the replica symmetric (RS)
solution $\phi_{\rm RS}(n)$ as 
$\phi_{\rm 1RSB}(n)=\mathop{\rm Extr}_{x} \{(n/x) \phi_{\rm RS}(x) \}$, 
where $\mathop{\rm Extr}_{x} \{\cdots \}$ denotes 
the operation of extremization with respect to $x$, 
the functional forms of $R(f)$ and $\Sigma(f)$ are 
in agreement \cite{Naka,Obuc}. 
In addition, the model class for which this property holds is rather wide, and
includes random energy models \cite{Derr,Gard2} 
and $p$-body spin glass models without external fields \cite{Naka}. 
This naturally motivates us to further explore more general relationships among 
$R(f)$, $\Sigma(f)$, and $\phi(n)$, including cases for which 
the formal accordance of functional forms between $R(f)$ and $\Sigma(f)$ does not hold. 

As a concrete effort for the exploration, 
we herein consider Ising perceptrons that store random input-output
patterns. There are two reasons for considering this system. 
First, the Ising perceptrons can be macroscopically characterized 
by a few sets of order parameters and are much easier to handle than 
systems of sparse couplings \cite{Naka2,Mona2,Meza1,Mont1,Zdeb}, 
for which several numerical calculations are required. 
Despite the simplicity, this model still could exhibit rich behavior 
in the phase space involving nontrivial RSB phenomena \cite{Gard,Krau}, 
which is highly suitable for our purpose. 
The second reason is that the meaning of complexity 
for the perceptrons of finite size 
is rather clear.
For the Ising perceptrons, 
a pure state at zero temperature can be identified with a stable cluster, 
the definition of which will be given in section \ref{sec:numerical}, 
with respect to single spin flips \cite{Cocc,Biro1,Arde}. 
For samples of small systems, the size of the clusters can be numerically 
evaluated by exhaustive enumeration without any ambiguity. 
This property is extremely useful for justifying theoretical predictions 
through numerical experiments.  

The remainder of the present paper is organized as follows. 
In the next section, we introduce the model that considered herein. 
In section 3, we provide a formalism that assesses the complexity and 
rate function based on the 1RSB evaluation of the 
generating function, for the Ising perceptrons. 
In the formalism, the complexity and rate function are defined not for 
the free energy $f$ but for the entropy $s$,
because the analysis is carried out for the micro-canonical ensemble 
of Ising weights that are perfectly compatible with a given set 
of random patterns. In section 4, we analyze the behavior of 
the weight space of the Ising perceptron using this formalism. 
It is found that for $\alpha=M/N < \alpha_s=0.833\ldots$, 
where $M$ is the number of random patterns, the typical phase space 
of the Ising perceptron is characterized by 
a {\em convex downward} complexity 
being equally dominated by a single large cluster of exponentially 
many weights and exponentially many small clusters of a single weight. 
For $\alpha > \alpha_s$, on the other hand, 
the rate function becomes relevant 
for the analysis because random patterns that are perfectly separable 
by the Ising weights are generated only atypically in this region. 
It is also found that a certain transition 
of 
the rate function occurs at another critical ratio $\alpha_{\rm GD}
=1.245\ldots$. These predictions are validated 
by comparison with the results of extensive numerical experiments 
in section \ref{sec:numerical}. The final section is devoted to a summary.

\section{Model definition}\label{sec:formulation}
A simple perceptron is a map from $\mR^N$ to $\{+1,-1\}$ defined as
\begin{equation}
y
= \left \{
\begin{array}{ll}
+1, & {\V{S}\cdot \V{x}}/{\sqrt{N}} > 0, \cr
-1, & {\V{S}\cdot \V{x}}/{\sqrt{N}} < 0, 
\end{array}
\label{eq:perceptron}
\right . 
\end{equation}
where $\V{x} \in \mR^N $ is the input pattern and $y \in \{+1,-1\}$ 
is the output label. The vector $\V{S}$ denotes the adjustable 
synaptic weight. We hereinafter focus on the case of Ising weight 
$S_{i} \in \{+1,-1\}$.
In a general scenario, the perceptron stores a given set of $M$ 
labeled patterns 
\begin{equation}
D^{M}=\{(\V{x}_{1},y_{1}),\cdots,(\V{x}_{M},y_{M})\}, \label{eq:examples}
\end{equation}
by adjusting the weight $\V{S}$ so as to completely reproduce 
the given label $y_\mu$ for the input $\V{x}_\mu$ for $\mu=1,2,\ldots,M$.

In the following, we consider the situation in which 
the patterns are independently and identically distributed 
samples from 
\begin{eqnarray}
P(\V{x})=\left( \frac{1}{\sqrt{2\pi}} \right)^N \exp \left(-\frac{\V{x}^2}{2} \right),\\
P(y)=\frac{1}{2}\left( \delta(y-1)+\delta(y+1) \right).
\end{eqnarray}
The question we address herein 
is how the space of the weights that store $D^{M}$ 
is characterized macroscopically when pattern ratio 
$\alpha=M/N \sim O(1)$ is fixed as $M$ and $N$ tend to infinity.

\section{Formalism}\label{sec:formalism}
\subsection{RS and 1RSB solutions of the generating function } 
As bases of our analysis, we first provide expressions of 
RS and 1RSB solutions of the generating function. Since these solutions have 
been derived numerous times in earlier studies \cite{Gard,Krau}, we present only a sketch of 
the derivation in the main text, and details are shown in
\ref{sec:Deri}. 
For readers who are not familiar with the replica method, we refer to
 \cite{STAT,INFO}.

We first define the Boltzmann factor 
$\eta(\V{S}|D^M)$ of the present system as
it takes $1$ if the weight $\V{S}$ is compatible with $D^M$ and $0$ 
otherwise\footnote{This is equivalent to the zero temperature 
limit $\beta \to \infty$ of the Boltzmann factor $e^{-\beta H(\V{S}|D^M)}$ 
where the Hamiltonian
$H(\V{S}|D^{M})$ is given by 
$\sum_{\mu=1}^{M}\Theta \left(
-y_{\mu}\frac{\V{S}\cdot\V{x}_{\mu}}{\sqrt{N}}
\right)$, which is equal to the number of patterns that are incompatible 
with the weight $\V{S}$.
 }. The explicit form is expressed as
\begin{equation}
\eta(\V{S}|D^M)
=\prod_{\mu=1}^M 
\Theta \left (-y_{\mu}\frac{\V{S}\cdot\V{x}_{\mu}}{\sqrt{N}}
\right),\label{eq:BF}
\end{equation}
where $\Theta(u)=1$ for $u > 0$ and $\Theta(u)=0$, otherwise. 
The partition function $Z(D^M)\equiv \sum_{\V{S}}
\eta(\V{S}|D^M)$ 
is equal to the number of weights 
 that are perfectly compatible with $D^M$ 
 in this situation, and 
varies randomly depending on the quenched randomness
$D^M$. 
This naturally leads us to evaluate
the generating function 
$\phi(n)=(1/N) \log \left [Z^n(D^M) \right ]_{D^M}$ 
using the replica method, 
where $\left [ \cdots \right ]_{D^M}$ represents the operation of 
averaging with respect to $D^M$. 
For $n=1,2,\ldots \in \mN$, this yields
the following expression: 
\begin{eqnarray}
\phi(n)&=&\mathop{\rm Extr}_{q^{ab},\widehat{q}^{ab}}
\left \{-\sum_{a <b}\widehat{q}^{ab}q^{ab}+ \log 
\left (
\sum_{S^1,S^2,\ldots,S^n} e^{\sum_{a<b}\widehat{q}^{ab} S^a S^b}  \right ) 
\right .\cr
&\phantom{=}&
\left . 
+\alpha \log \left [ \prod_{a=1}^n \Theta(u^a) \right ]_{\V{u}} \right \}, 
\label{replica_int}
\end{eqnarray}
where $[\cdots]_{\V{u}}$ represents averaging with respect to 
multivariate Gaussian random variables $u^1,u^2,\ldots,u^n$, the first 
and second moments of which are specified as 
$\left [u^a \right ]_{\V{u}}=0$ and $\left[ u^a u^b \right ]_{\V{u}}=
\delta_{ab}+(1-\delta_{ab})q^{ab} $ $(a,b=1,2,\ldots,n)$, respectively. 

Analytical continuation from $n \in \mN$ to $n \in \mR$ is performed
by imposing a certain permutation symmetry on the extremum point 
of the right-hand side of (\ref{replica_int}). 
We find several solutions in the RS and 1RSB levels. 

\subsubsection{RS solutions}\label{sec:RSsol}
Constraints $q^{ab}=q$ and $\widehat{q}^{ab}=\widehat{q}$ characterize
the RS solutions. Solving the extremization 
problem of (\ref{replica_int})
analytically and numerically under these constraints yields the following 
two solutions: \\
\noindent{\bf RS1:} $0<q<1$ and $\widehat{q} < + \infty$. 
\begin{eqnarray}
\phi_{\rm RS1}(n)&=&
 -\frac{n(n-1)}{2}q\Wh{q}-\frac{1}{2}n\Wh{q} + \log  
\left (\int Dz
\left ( 2 \cosh 
\left (\sqrt{\widehat{q}} z 
\right )
\right )^n
\right ) \nonumber \\
&&+\alpha \log  \left (\int Dz
E^n\left (\sqrt{\frac{q}{1-q}}z \right ) \right ), 
\label{RS1}
\end{eqnarray}
where $Dz=\exp\left (-z^2/2 \right )/\sqrt{2 \pi}$ represents the Gaussian 
measure and $E(u)=\int_u^{+\infty} Dz$. \\
\noindent{\bf RS2:} $q=1$ and $\widehat{q} = + \infty$. 
\begin{eqnarray}
\phi_{\rm RS2}(n)=(1-\alpha) \log 2. 
\label{RS2}
\end{eqnarray}

\subsubsection{1RSB solutions}\label{sec:1RSBsol}
In 1RSB solutions, replica indices are divided into $n/m$ groups
of identical size $m$. Constraints for characterizing the 1RSB solutions
are expressed as 
\begin{eqnarray}
q^{ab}=\left \{
\begin{array}{ll}
q_1 & \mbox{if $a$ and $b$ belong to the same group}, \cr
q_0 & \mbox{otherwise},
\end{array}
\right .
\label{1RSB_const}
\end{eqnarray}
and are similarly expressed for $\widehat{q}^{ab}$. 
Three solutions are found under these constraints: \\
\noindent{\bf 1RSB1:} $(q_1,q_0)=(1,q)$ and $(\widehat{q}_1, \widehat{q}_0)
=(+\infty, \widehat{q})$, where $q$ and $\widehat{q}$ take the same values as those for
$\phi_{\rm RS1}(n)$. 
\begin{eqnarray}
\phi_{\rm 1RSB1}(n,m)=\phi_{\rm RS1}\left (\frac{n}{m} \right ). 
\label{1RSB1}
\end{eqnarray}

\noindent{\bf 1RSB2:} $(q_1,q_0)=(q,q)$ and $(\widehat{q}_1, \widehat{q}_0)
=(\widehat{q},\widehat{q} )$, where $q$ and $\widehat{q}$ take the same values as those for 
$\phi_{\rm RS1}(n)$.
\begin{eqnarray}
\phi_{\rm 1RSB2}(n,m)=\phi_{\rm RS1}(n). 
\label{1RSB2}
\end{eqnarray}

\noindent{\bf 1RSB3:} $(q_1,q_0)=(1,1)$ and $(\widehat{q}_1, \widehat{q}_0)
=(+\infty, +\infty)$. 
\begin{eqnarray}
\phi_{\rm 1RSB3}(n,m)=\phi_{\rm RS2}(n)=(1-\alpha)\log 2. 
\label{1RSB3}
\end{eqnarray}

In usual analyses, Parisi's 1RSB parameter $m$ is determined 
by the extremum condition in evaluating 
$\phi(n)=\mathop{\rm Extr}_{m}\left \{\phi_{\rm 1RSB*}(n,m) \right \}$, 
where $\rm *=1,2$ and $\rm 3$. In addition, there might be no need
to classify ${\bf 1RSB2}$ and ${\bf 1RSB3}$ as 
1RSB solutions because ${\bf 1RSB2}$ and ${\bf 1RSB3}$ are completely reduced to 
${\bf RS1}$ and ${\bf RS2}$, respectively. However, handling these 
three solutions as 1RSB solutions, leaving the $m$-dependence 
of $\phi_{\rm 1RSB}(n,m)$ explicitly, is crucial
for the current purpose of relating the concepts of 
$\phi(n)$, $\Sigma(s)$, and $R(s)$ based on physical considerations 
presented in the following subsection. 

\subsection{1RSB solution as the generating function of complexity 
and rate function}
Let us present the 1RSB solutions through a physical inference 
based on arguments presented in earlier 
studies \cite{Mona1,Mona3,Mont1,Cocc,Biro1}. 
Here, we assume a situation in which the weight space is divided into 
exponentially many 
pure states for a given sample of $D^M$. 

We introduce an indicator function 
$\delta_\gamma(\V{S})$, which is defined as
$\delta_\gamma(\V{S})=1$, if $\V{S}$ belongs to the pure state 
$\gamma$ and $0$, otherwise, to express the number of 
weights included in $\gamma$ as
\begin{eqnarray}
Z_\gamma=\sum_{\V{S}} 
\prod_{\mu=1}^M 
\Theta \left (-y_{\mu}\frac{\V{S}\cdot\V{x}_{\mu}}{\sqrt{N}}
\right) \delta_\gamma(\V{S}). 
\label{pure_state_volume}
\end{eqnarray}

Let us assume that $Z_\gamma$ typically scales as
$Z_\gamma \sim \exp (Ns)$, where $s \sim O(1)$ has the physical 
meaning of entropy (density), and 
the number of pure states corresponding to the value of the entropy 
$s$ increases as ${\cal N}(s) \sim \exp (N\Sigma(s) )$, where
$\Sigma(s) \ge 0$ is the complexity for the entropy $s$. 
This assumption, in conjunction with the saddle point 
assessment, provides us with a generating function of 
$\Sigma(s)$, as follows:
\begin{eqnarray}
g(x|D^M)&=&\frac{1}{N}\log 
\left (\sum_\gamma Z_\gamma^x \right)
\sim \frac{1}{N} \log \left (
\int ds e^{N(xs+\Sigma(s))}\right ) \cr
&=&\mathop{\rm max}_{s}\left \{ xs + \Sigma(s) \right \}.
\label{generating_sigma}
\end{eqnarray}
This relationship indicates that when $\Sigma(s)$ is a convex upward function, 
it can be assessed from $g(x|D^M)$ as
\begin{eqnarray}
s(x)=\frac{\partial g(x|D^M)}{\partial x}, \quad
\Sigma(s(x))=g(x|D^M)-x \frac{\partial g(x|D^M)}{\partial x}, 
\label{complexity_Legendre}
\end{eqnarray}
being parameterized by $x$. 
Here, $g(x|D^M)$ is defined for each sample of $D^M$. 
However, the self-averaging property is assumed to hold
in the current system, which means that 
$g(x|D^M)$ for typical samples converges to its
average $g(x)=\left [g(x|D^M) \right ]_{D^M}$ 
in a large system limit of 
$N,M \to \infty$ while maintaining $\alpha=M/N \sim O(1)$. 

The replica method can be used to assess $g(x)$. 
For this, we consider the following identity:
\begin{eqnarray}
g(x)=\frac{1}{N} \left [ \log \left (\sum_\gamma Z_\gamma^x \right )
\right ]_{D^M} 
= \lim_{y \to 0}
\frac{\partial }{\partial y}\left (\frac{1}{N} 
\log \left [ \left (\sum_\gamma Z_\gamma^x \right )^y \right ]_{D^M} 
\right ).
\label{replica_trick_g}
\end{eqnarray}
Although exact evaluation of the right-hand side of (\ref{replica_trick_g})
is difficult, for $x,y \in \mN$, the equation (\ref{pure_state_volume}) and
the formula of series expansion provide the following expression:
\begin{eqnarray}
&&\left [ \left (\sum_\gamma Z_\gamma^x \right )^y \right ]_{D^M}  \cr
&&=\sum_{\{\gamma^\sigma\}}\sum_{\{\V{S}^{\sigma a} \}}
\left [\prod_{\mu=1}^M 
\prod_{\sigma=1}^y \prod_{a=1}^x 
\Theta \left (-y_{\mu}\frac{\V{S}^{\sigma a}\cdot\V{x}_{\mu}}{\sqrt{N}}
\right) \right ]_{D^M}
\prod_{\sigma=1}^y \prod_{a=1}^x 
\delta_{\gamma^\sigma}(\V{S}^{\sigma a}), 
\label{g_1RSB}
\end{eqnarray}
which can be evaluated by the saddle point method in the large system limit. 

\begin{figure}[htbp]
\begin{center}
   \includegraphics[height=50mm,width=60mm]{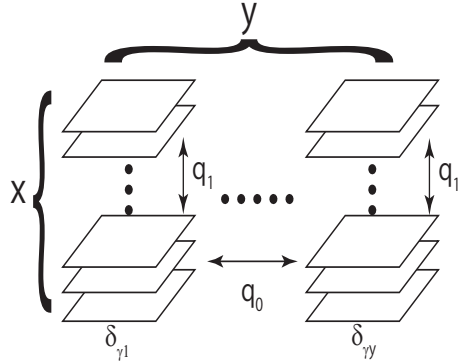}
 \caption{Schematic diagram of the 1RSB structure 
of the factor $\prod_{\sigma=1}^{y}
\prod_{a=1}^{x}\delta_{\gamma^{\sigma}}(S^{\sigma a})$.} 
 \label{fig:1RSB}
\end{center}
\end{figure}

The following observations are noteworthy in the evaluation. 
\begin{itemize}
\item The summation is taken over all possible configurations 
of $xy$ replica weights. 
\item However, the factor of $\prod_{\sigma=1}^y \prod_{a=1}^x 
\delta_{\gamma^\sigma}(\V{S}^{\sigma a})$
allows only contributions from configurations in which $xy$ replica weights 
are equally assigned to $y$ pure states by $x$. 
\end{itemize}
These observations are nothing more than the physical meaning of the 1RSB ansatz
in assessing $\left [Z^n(D^M) \right ]_{D^M}$
with substitution of $n=xy$ and $m=x$
(figure \ref{fig:1RSB}).
Accepting this interpretation yields the following expression:
\begin{eqnarray}
\frac{1}{N} 
\log \left [ \left (\sum_\gamma Z_\gamma^x \right )^y \right ]_{D^M} 
=\phi_{\rm 1RSB}(xy,x), 
\label{phi1RSB}
\end{eqnarray}
where $\phi_{\rm 1RSB}(n,m)$ is the 1RSB solution considered in the previous 
section, and its concrete functional form should be chosen appropriately from among
{\bf 1RSB1}, {\bf 1RSB2}, and {\bf 1RSB3} for a given pair of $\alpha$ and $n$. 

Inserting (\ref{phi1RSB}) into (\ref{replica_trick_g})
yields $g(x)=x (\partial/\partial n) \phi_{\rm 1RSB}(n,x) |_{n=0}$,
which directly yields the following formula relating typical complexity 
to $\phi_{\rm 1RSB}(n,m)$:
\begin{eqnarray}
\left \{
\begin{array}{l}
s(x)=(\partial /\partial x)
\left (x ({\partial}/{\partial n})
\phi_{\rm 1RSB}(n,x) |_{n=0} \right ),  \cr
\Sigma(s(x))=
- x^2({\partial}^2/{\partial x} \partial n)
\phi_{\rm 1RSB}(n,x) |_{n=0}. 
\end{array}
\right . 
\label{phi_complexity}
\end{eqnarray}
On the other hand, 
an identity with respect to the indicator function 
$\sum_{\gamma} \delta_\gamma(\V{S})=1$ for $\forall{\bm{S}}$
guarantees $\sum_{\gamma}Z_\gamma=Z(D^M)$, indicating 
that $\phi(n)=\phi_{\rm 1RSB}(n,x)|_{x=1}$ holds in general. 
This means that the rate function can be assessed from $\phi_{\rm 1RSB}(n,m)$ 
as follows:
\begin{eqnarray}
\left \{
\begin{array}{l}
s_{\rm tot}(n)=(\partial /\partial n)
\phi_{\rm 1RSB}(n,x) |_{x=1},  \cr
R(s_{\rm tot}(n))=-n^2(\partial/\partial n)
\left ( n^{-1} \phi_{\rm 1RSB}(n,x) |_{x=1} \right ),
\end{array}
\right. 
\label{phi_rate}
\end{eqnarray}
where we define the total entropy 
$s_{\rm tot}
\equiv\lim_{N\to \infty}(1/N)\log Z
=\max_{s}\{s+\Sigma(s) \}
$
which corresponds to 
the total number of weights that are compatible with $D^{M}$.
In (\ref{phi_complexity}), the parameter $x$ can vary only in such a range 
that both $s(x) \ge 0$ and $\Sigma(s(x)) \ge 0$ hold. 
Similarly, the conditions $s_{\rm tot}(n) \ge 0$ and 
$R(s_{\rm tot}(n)) \le 0$ restrict
the range of $n$ in (\ref{phi_rate}). 
These constitute the main result of the present paper. 

Here, three issues are noteworthy. First, for a class of disordered systems, 
including random energy models and $p$-body spin glass models
without external fields, 
two equalities
$\phi_{\rm 1RSB}(n,m)=(n/m) \phi_{\rm RS}(m)$ 
and $\phi_{\rm 1RSB}(n,m=1)=\phi_{\rm RS}(n)$, hold 
in assessing the complexity and rate function,
respectively,  where $\phi_{\rm RS}(n)$ is an identical 
RS solution of the generating function $\phi(n)$. 
Inserting these functions into (\ref{phi_complexity}) and 
(\ref{phi_rate}) offers an identical functional form for 
both the complexity and the rate function, while their 
domains of definition are disjointed, except for a point of the typical value of 
free energy $f^*$ (or entropy $s^*$). 
The current system, however, does not possess 
this property because $\phi_{\rm 1RSB}(n,m)=(n/m) \phi_{\rm RS}(m)$ 
does not hold for {\bf 1RSB1}, {\bf 1RSB2}, or {\bf 1RSB3}
while $\phi_{\rm 1RSB}(n,m=1)=\phi_{\rm RS}(n)$ is always satisfied.
Second, (\ref{phi_complexity}) and (\ref{phi_rate}) are 
valid only when $\phi_{\rm 1RSB}(n,m)$ are stable against any perturbation 
for a further RSB. Fortunately, in the present problem, 
a stable solution against any known 
RSB instabilities can be constructed for $\forall{\alpha} > 0$ and $\forall{n}>0$.   
This implies that, in the present analysis, there is no need to consider further RSB. 
Finally, however, we have to keep in mind that (\ref{phi_complexity}) and (\ref{phi_rate}) depend on the 
assumptions that correct $\Sigma(s)$ and $R(s)$ are convex upward
functions, respectively. 
When the convex upward property does not hold, the estimates of 
(\ref{phi_complexity}) and (\ref{phi_rate}) 
represent not the correct solution, but rather its convex hull. 
The following analytical and experimental assessment indicates
that this is the case for $\Sigma(s)$ of sufficiently low $\alpha$
and $R(s)$ of sufficiently high $\alpha$.

\section{Theoretical predictions}
We are now ready to use the formalism developed above
to analyze the behavior of the weight space of the Ising perceptron.

\subsection{Complexity for $\alpha< \alpha_{\rm s}=0.833 \ldots$}
In order to perform the analysis, it is necessary to select 
a certain solution (functional form) from among the three candidates 
of {\bf 1RSB1}, {\bf 1RSB2}, and {\bf 1RSB3}. 
Analyticity and physical plausibility are two guidelines for this task. 

The replica method is a scheme to infer the properties for real replica 
numbers $n \in \mR$ by analytical continuation 
from those for natural numbers $n = 1,2,\ldots \in \mN$. 
This indicates that, for examining typical ($n\to 0$) behavior, 
it is plausible to select the solution 
of $\phi(n)$ that is dominant around $n \ge 1$, because 
unity is the natural number that is closest to zero. 
For $\alpha < \alpha_{\rm s}=0.833 \ldots$, this solution
is $\phi_{\rm RS1}(n)$. 
In addition, the relevant $\phi_{\rm 1RSB}(n,m)$ must agree with this 
solution at $m=1$. 
These considerations offer two candidates of $g(x)$ as
\begin{eqnarray}
g_{\rm 1RSB1}(x)=
x \frac{\partial}{\partial n}\phi_{\rm 1RSB1}(n,x)|_{n=0}
=\phi_{\rm RS1}^\prime(0), 
\label{g1RSB1}
\end{eqnarray}
and 
\begin{eqnarray}
g_{\rm 1RSB2}(x)=
x \frac{\partial}{\partial n}\phi_{\rm 1RSB2}(n,x)|_{n=0}
=x \phi_{\rm RS1}^\prime(0). 
\label{g1RSB2}
\end{eqnarray}

We combine these solutions to construct an entire functional 
form of $g(x)$ based on physical considerations. 
For $x \gg 1$, $g(x)$ should vary approximately linearly with respect to 
$x$, because a single pure state of the largest entropy 
typically dominates $\sum_\gamma Z_\gamma^x$. 
In addition, $s(x)=(\partial/\partial x) g(x)$ for $x \sim 0$
should be smaller than that for $x \gg 1$ because $s(x)$ should increase
monotonically with respect to $x$. 
Furthermore, $g(x)$ must be a continuous function. These considerations reasonably 
yield an entire functional form of $g(x)$ as
\begin{eqnarray}
g(x)=\left \{
\begin{array}{ll}
\phi_{\rm RS1}^\prime(0), & x \leq 1,\\
x\phi_{\rm RS1}^\prime(0), & x > 1,
\end{array}
\right .
\label{gx}
\end{eqnarray}
which yields the complexity as
\begin{eqnarray}
\Sigma(s)=\left \{
\begin{array}{ll}
\phi_{\rm RS1}^\prime(0)-s, & 0\leq s \leq \phi_{\rm RS1}^\prime(0), \\
-\infty, & \mbox{otherwise}.
\end{array}
\right .
\label{typicalcomplexity}
\end{eqnarray}

\begin{figure}[htbp]
\begin{center}
   \includegraphics[height=50mm,width=50mm]{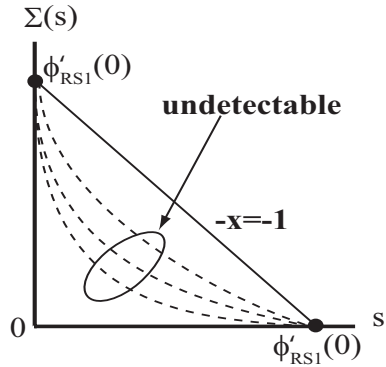}
 \caption{Schematic profile of complexity (\ref{typicalcomplexity}). 
The characteristic exponent of the size distribution of
pure states cannot be correctly assessed in the current formalism if 
it is a convex downward function (dashed curves). 
In such cases, the complexity $\Sigma(s)$ assessed from $g(x)$ (solid line)
is the convex hull (black circle) of the correct exponent. 
}
\label{fig:typ-g(x)}
\end{center}
\end{figure}

The piecewise linear profile of (\ref{gx}) is 
somewhat extraordinary.
This is thought to be because the correct complexity 
is not convex upward in this system. 
When $\Sigma(s)$ is convex upward, 
the current formalism using the saddle-point method 
defines a one-to-one map between 
$g(x)$ and $\Sigma(s)$. However, if $\Sigma(s)$ is not
convex upward, the functional profile of a region in which  
the correct complexity is convex downward is lost and 
only the convex hull is obtained by the transformation from $g(x)$, 
as shown in figure \ref{fig:typ-g(x)}. 
The piecewise liner profile of $g(x)$ presumably signals that 
this actually occurs in the current problem. 
Similar behavior of the complexity could also be observed 
in a certain type of random energy models \cite{Bouc}.

The physical implication of (\ref{typicalcomplexity}), 
the profile of which is obtained by connecting
two points $(s,\Sigma)=(0,\phi_{\rm RS1}^\prime(0))$ 
and $(\phi_{\rm RS1}^\prime(0), 0)$ with a straight line 
having a slope of $-x=-1$, is that the weight space is equally 
dominated by exponentially many clusters of vanishing entropy 
and a subexponential number of large clusters composed 
of exponentially many weights. 
The existence of large clusters may accord with 
an earlier study which reported that 
local search heuristics of a certain type 
manage to find a compatible weight efficiently
up to a considerably large value of $\alpha$ near to the 
capacity $\alpha_{s}$ \cite{Brau}.
On the other hand, the coexisting exponentially many 
small clusters may be a major origin of a known difficulty 
in finding compatible weights by Monte Carlo sampling schemes
\cite{Horner1,Horner2}. 

\subsection{Rate function for $\alpha> \alpha_{\rm s}=0.833 \ldots$\label{sec:typ} 
and a transition at $\alpha_{\rm GD}=1.245 \ldots$}
For $\alpha_{\rm s} < \alpha$, (\ref{typicalcomplexity}) becomes 
negative, which implies that there exist no compatible weights for 
typical samples of $D^M$. In such cases, the rate function $R(s)$, 
which characterizes a small probability that atypical samples 
that are compatible with the Ising perceptrons are generated, 
becomes relevant in the current analysis. Therefore, 
we focus on the assessment of this exponent for this region. 

\begin{figure}[htbp]
\hspace{-7mm}
\begin{minipage}{0.32\hsize}
\begin{center}
   \includegraphics[height=40mm,width=50mm]{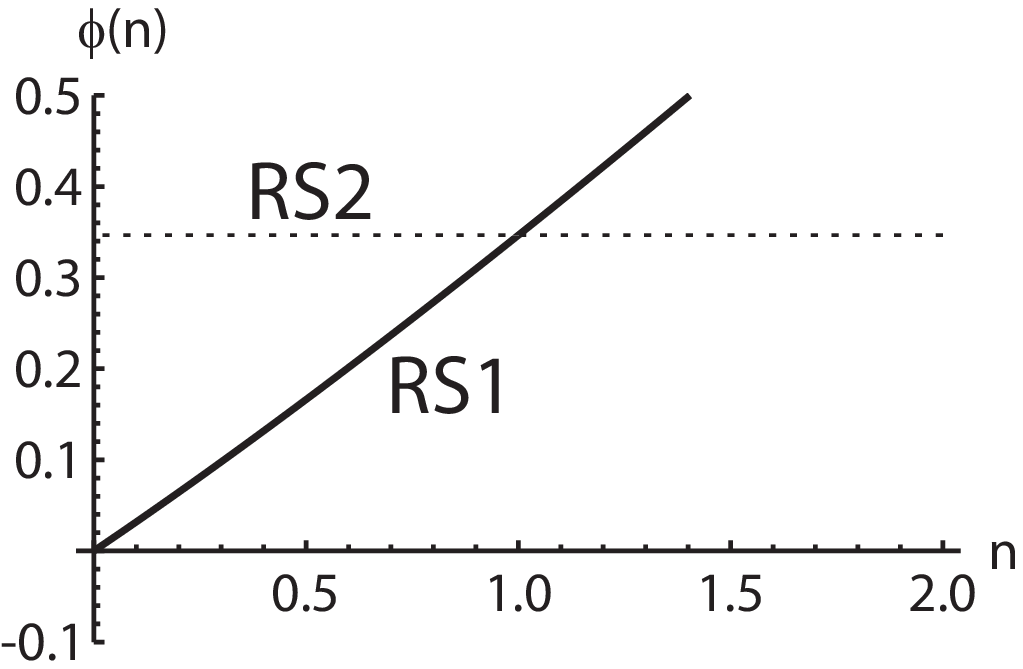}
\end{center}
\end{minipage}
\hspace{2mm}
 \begin{minipage}{0.32\hsize}
\begin{center}
   \includegraphics[height=40mm,width=50mm]{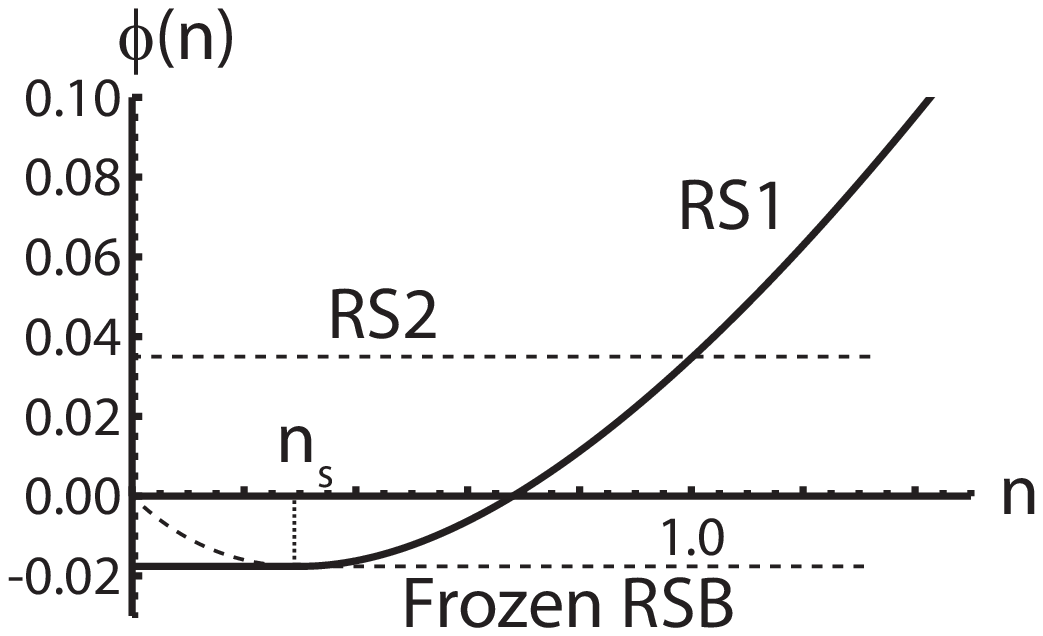}
\end{center}
 \end{minipage}
\hspace{2mm}
 \begin{minipage}{0.32\hsize}
\begin{center}
   \includegraphics[height=40mm,width=50mm]{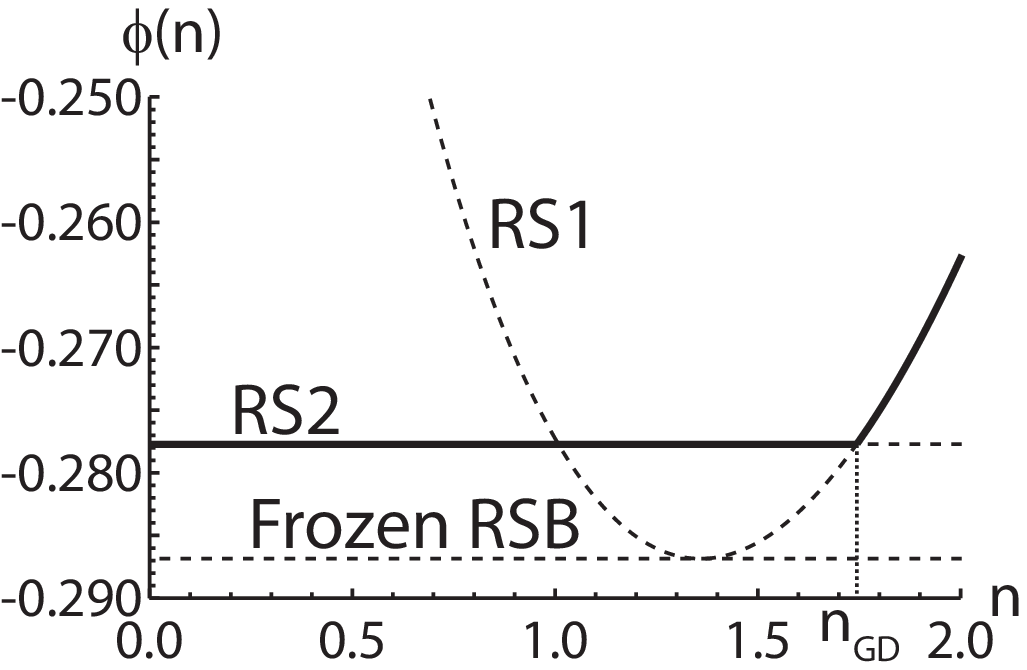}
\end{center}
 \end{minipage}
\caption{Behavior of $\phi(n)$. The solid lines denote 
the correct $\phi(n)$, and the 
dotted lines are the RS and frozen RSB branches. 
The corresponding values of 
the parameter $\alpha$ are $0.5,0.95$, and 
$1.4$, from left to right.}
\label{fig:phi2}
\end{figure}

For $\alpha_{\rm s} < \alpha < \alpha_{\rm GD}=1.245\ldots$, 
$\phi_{\rm RS1}(n)$ dominates the generating function $\phi(n)$
in the vicinity of $n \ge 1$ as for $\alpha < \alpha_{\rm s}$. 
This means that $\phi_{\rm 1RSB1}(n,m=1)=\phi_{\rm 1RSB2}(n,m=1)=
\phi_{\rm RS1}(n)$ should be used to assess $R(s)$ of 
relatively frequent events that correspond to $0<n<1$. 
However, this function is minimized to a negative value 
at a certain point at which $0<n_{\rm s}(\alpha)<1$, which implies
that assessment by na\"{i}vely using 
$\phi_{\rm RS1}(n)$ for $n<n_{\rm s}(\alpha)$ 
leads to incorrect results, which yield 
a negative total entropy $s_{\rm tot}(n)
=(\partial/\partial n)\phi_{\rm RS}(n) < 0$. 
In order to avoid this inconsistency, we fix the value of
$\phi(n)$ to $\phi_{\rm RS1}(n_{\rm s}(\alpha))$, which is reduced to 
the conventional construction of a frozen RSB solution. 
In particular, this yields an assessment of 
\begin{eqnarray}
R(0)=\phi_{\rm RS1}(n_{\rm s}(\alpha))=
\mathop{\rm min}_{n} \{\phi_{\rm RS1}(n) \}, 
\label{separable_rate}
\end{eqnarray}
which has the physical meaning of a characteristic exponent of 
a small probability that a given sample set $D^M$ is 
separable by certain Ising perceptrons. 
For $\alpha \geq \alpha_{\rm GD}=1.245 \ldots$, 
on the other hand, 
the dominant solution of $\phi(n)$ in the vicinity of $n \ge 1$
is updated from $\phi_{\rm RS1}(n)$ to $\phi_{\rm RS2}(n)
=(1-\alpha)\log 2$, which yields 
\begin{eqnarray}
R(0)=\phi_{\rm RS2}(n)=(1-\alpha)\log 2. 
\label{separable_rate2}
\end{eqnarray}
In order to provide a visual representation of the above discussions, 
we depict the behaviors of $\phi(n)$ in figure \ref{fig:phi2}.

The difference in physical behavior 
between $\alpha_{\rm s} < \alpha < \alpha_{\rm GD}$ 
and $\alpha > \alpha_{\rm GD}$ is expected to be as follows. 
For $\alpha_{\rm s} < \alpha < \alpha_{\rm GD}$, the dominant solution  
around $n>n_{\rm s}(\alpha)$, $\phi(n)=\phi_{\rm RS1}(n)$, varies smoothly.
This leads to the following behavior of $R(s)$ in the vicinity of $s =0$: 
\begin{eqnarray}
R(s) = R(0) -A  s^2 +\ldots, 
\label{below_alpha_s}
\end{eqnarray}
where $A >0$ is a certain constant, 
which implies that large clusters can appear with a relatively 
large probability although typical samples of $D^M$ are not
separable by the Ising perceptrons. 
On the other hand, for $\alpha > \alpha_{\rm GD}$, 
$\phi(n)=\phi_{\rm RS2}(n)$ is constant for $n<n_{\rm GD}(\alpha)$,
which is characterized by $\phi_{\rm RS2}(n_{\rm GD}(\alpha))=
\phi_{\rm RS1}(n_{\rm GD}(\alpha))$ 
and $n_{\rm GD}(\alpha)>1$, and is switched to $\phi(n)=\phi_{\rm RS1}(n)$ 
for $n>n_{\rm GD}(\alpha)$ at $n=n_{\rm GD}(\alpha)$, which is accompanied by 
a jump in the first derivative. This indicates that 
(upward) 
convexity does not
hold for $R(s)$ in the region of $0 < s < (\partial /\partial n) 
\phi_{\rm RS1}(n_{\rm GD}(\alpha))$ as was mentioned for $\Sigma(s)$
in the previous subsection, which implies that the events of 
$s=0$ overwhelm those of $s>0$ in relative probabilities. 
Therefore, the generation of large clusters should be considerably 
rare for $\alpha$ of this region.

\subsection{Phase diagram on the $n$-$\alpha$ plane}
\begin{figure}[htbp]
\begin{center}
   \includegraphics[height=50mm,width=60mm]{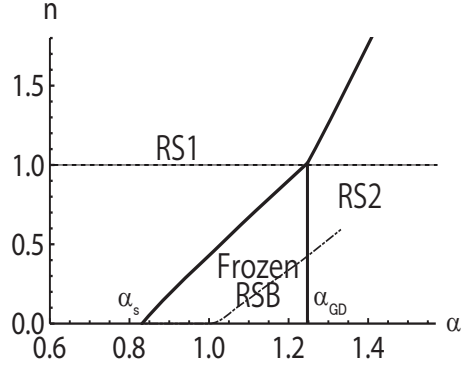}
 \caption{Phase diagram on the $n$-$\alpha$ plane. 
Solid lines are phase boundaries, and the dotted line denotes 
$n=1$. The dotted-dashed line expresses the AT line for the RS1 
solution, but is irrelevant. 
The AT line vanishes at a certain value of $\alpha$, 
because the solution for $0<q<1$ vanishes at this point.
The RS2 solution involves the AT instability only on the $n=0$ line, 
which is presumably of no relevance in the replica analysis. 
}
 \label{fig:PD}
\end{center}
\end{figure}

The above considerations are sufficient to draw a phase diagram 
on the $n$-$\alpha$ plane, which is depicted in figure \ref{fig:PD}. 

The value of the tricritical point 
$\alpha_{\rm GD}=1.245 \ldots$ is identical to the critical 
ratio of the perfect learning of the Ising perceptrons in the teacher-student 
scenario \cite{Gyor,Enge}. 
Formally, this agreement is explained as follows. 
The dominant solution for $n < 1$ is determined by 
whether $\phi_{\rm RS1}(n)$ or $\phi_{\rm RS2}(n)$ dominates
around $n \ge 1$. Since $\phi_{\rm RS1}(n=1)=\phi_{\rm RS2}(n=1)$
is always guaranteed, the critical condition is 
given as $(\partial /\partial n)\phi_{\rm RS1}(n)|_{n=1}=
(\partial /\partial n)\phi_{\rm RS2}(n)|_{n=1}=0$. 
On the other hand, $(\partial /\partial n)\phi_{\rm RS1}(n)|_{n=1}$ 
generally provides the total entropy after learning in the teacher-student scenario, 
the target of which can be dealt with as an $(n+1)$-replicated system, in which 
the teacher is handled as an extra replica. 
Therefore, the condition of perfect learning, which 
indicates that the weight of the student agrees perfectly 
with that of the teacher after learning, 
is identical to the vanishing entropy condition of the 
$(n+1)$-replicated system in the limit $n\to 0$, which 
agrees with $(\partial /\partial n)\phi_{\rm RS1}(n)|_{n=1}=0$, 
giving the critical value $\alpha_{\rm GD}(\alpha)$ in the current problem. 
Although the agreement is justified formally in this manner, 
its physical implication remains somewhat unclear. 
The line $n=1$, which passes through the tricritical point, 
may have an analogous relation to the 
concept of Nishimori's line in the theory of spin glasses \cite{STAT,Nish}. 

Finally, we mention the de 
Almeida-Thouless (AT) condition in this model \cite{Alme}. 
The AT (stability) condition of $\phi_{\rm RS}(n)$ with the 
order parameters $q$ and $\Wh{q}$
is expressed as follows:
\begin{eqnarray}
\frac{\alpha}{(1-q)^2}
\frac{
\int Dz 
E^{n}
\left(
\frac{E^{\prime \prime}}{E}
-
\left(
\frac{E^\prime}{E}
\right)^{2}
\right)^2
}
{
\int Dz 
E^{n}
}
\frac{\int Dz \cosh^{n-4} \sqrt{\Wh{q}}z }{
\int Dz \cosh^n \sqrt{\Wh{q}}z
}
\leq 1.\label{eq:AT}
\end{eqnarray}
An outline of the derivation is given in \cite{Gard}. 
This condition for $\phi_{\rm RS1}(n)$ 
is broken in a certain region on the $n$-$\alpha$ 
plane, but is irrelevant because the region is always 
included in $n < n_{s}(\alpha)$, for which the relevant solution is 
already switched to that of the frozen RSB. 
On the other hand, $\phi_{\rm RS2}$ is stable for $n>0$ 
but becomes unstable only on $n=0$, as reported 
in \cite{Gard}.
The relevance of this instability for $\alpha\geq \alpha_{\rm GD}$ may 
require more a detailed discussion,
but we assume herein that this instability can be ignored 
because only the asymptotic behavior of $\phi(n)$ in the 
limit $n\to 0$ is relevant in procedures of the replica method.

\section{Numerical validation}\label{sec:numerical}
For validating the theoretical predictions obtained 
in the previous section, we carried out extensive numerical experiments. 
In describing the experiments, let us first define the cluster in the 
present problem. 
The cluster is a set of spin configurations that are stable
with respect to single spin flips \cite{Cocc,Biro1,Arde}. 
Clusters have the following properties:
\begin{itemize}
\item{Any configuration belongs to a cluster.}
\item{
When a spin configuration ``A'' can be moved to another 
configuration ``B'' by a single spin flip without 
changing the number of incompatible patterns, 
``A'' and ``B'' belong to the same cluster.
}
\end{itemize}
In the following, we concentrate on vanishing energy clusters, which are 
composed of weights that are perfectly compatible with $D^M$. 

Before going into details, 
we elucidate the relation between the cluster and the pure state.
Identifying the microscopic description of a pure state is generally 
a delicate problem, but
in the Ising perceptron 
a pure state  
can be identified with a cluster,
as mentioned in section \ref{sec:intro}. 
There is no proof of this statement but 
it is naturally understood by 
considering the following aspects of the present problem: 
The Boltzmann weight of 
$\eta(\V{S}|D^M)$ in (\ref{eq:BF})
becomes completely zero if there is any incompatible pattern. 
This means that accessing from a cluster to a different cluster 
by single spin flips 
is impossible because those clusters are completely separated by states
with zero probability 
$\eta(\V{S}|D^M)=0$.
This naturally leads to identifying a cluster with a pure state, 
because a pure state is a set of configurations 
which cannot be accessed from other sets by natural dynamics.
Several earlier studies support this description 
\cite{Cocc,Biro1,Arde}, 
and we hereafter admit this 
assumption.

Now, let us return to the experiments. We denote the size of a cluster 
as $Q$ and the number of size-$Q$ 
clusters for a sample $D^M$ as $C(Q|D^{M})$.  
the 
entropy of a cluster $s$ is considered to be identified by $s=(1/N)\log Q$, and the complexity
$\Sigma(s|D^{M})$ corresponds to $(1/N)\log C(Q|D^{M})$.
The clusters can be numerically evaluated, and hence 
we can construct the 1RSB generating function from the numerical data as
\begin{equation}
\phi_{\rm 1RSBnum}(n=xy,m=x)=\frac{1}{N}\log
\left[
\left(
\sum_{\gamma}Q_{\gamma}^{x}
\right)^y
\right],
\end{equation}
where $\left [ \ldots \right ]$ denotes the sample average operation
with respect to $D^M$. 
In the typical limit $y\to 0$, 
this yields the following expression:
\begin{equation}
g_{\rm num}(x)=
\lim_{y \to 0} \frac{\partial}{\partial y} \phi_{\rm 1RSBnum}(xy,x)
=
\frac{1}{N}
\frac{
\left[
\Theta
\left(
\sum_{\gamma}Q_{\gamma}^{x}
\right)
\log
\left(
\sum_{\gamma}Q_{\gamma}^{x}
\right)
\right]
}
{
\left[
\Theta
\left(
\sum_{\gamma}Q_{\gamma}^{x}
\right)
\right]
}
\label{eq:gnum},
\end{equation}
where the step function $\Theta(x)$ comes from the differentiation of 
$\log\left[
\left(
\sum_{\gamma}Q_{\gamma}^{x}
\right)^y
\right]$ with respect to $y$. 
This means that if there is no cluster for a sample $D^M$, 
then the contribution of $\Theta
\left(
\sum_{\gamma}Q_{\gamma}^{x}
\right)
\log
\left(
\sum_{\gamma}Q_{\gamma}^{x}
\right)$ vanishes.

In order to examine the consistency with the replica analysis, 
we assess (\ref{eq:gnum}) based on data obtained in extensive 
numerical experiments. The function $g_{\rm num}(x)$ is evaluated by the exact 
enumeration of weights that are compatible with $D^M$, which are 
referred to hereinafter as {\em solutions}. 
The procedure is summarized as follows: 
\begin{enumerate}
\item{Generate $M$ examples $D^{M}=\{
(y_{1},\V{x}_{1})
\cdots (y_{M},\V{x}_{M})
\}$.} 
\item{
Enumerate all solutions.
} 
\item{
Partition the solutions into clusters, and calculate
$\sum_{\gamma}Q_{\gamma}^{x}$ for an appropriate set of $x$. 
We actually took $41$ points between $x=0$ and $2.0$. 
}
\item{
Repeat the above procedures until sufficient data are obtained and 
calculate $
\left[
\Theta
\left(
\sum_{\gamma}Q_{\gamma}^{x}
\right)
\log
\left(
\sum_{\gamma}Q_{\gamma}^{x}
\right)
\right]
$ by taking the sample average.  
}
\end{enumerate} 

\begin{figure}[htbp]
\begin{tabular}{cc}
\hspace{-2mm}
\begin{minipage}[t]{0.48\hsize}
\begin{center}
 \includegraphics[height=50mm,width=60mm]{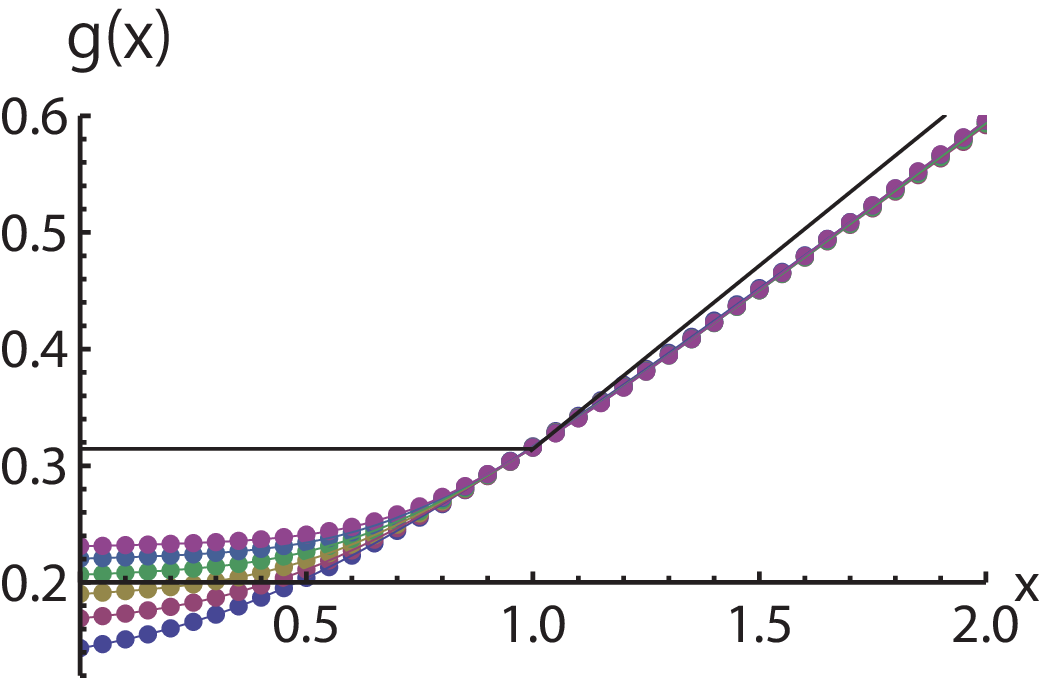}
 \caption{
Behavior of $g_{\rm num}(x)$
for $\alpha=0.5$. 
The system sizes are $N=14,16,18,20,22$, and $24$, 
from bottom to top. 
The solid lines denote $g(x)$ given by (\ref{gx}).
The number of samples is $32,000$ for each $N$. 
Error bars are smaller than the size of markers.
As the system size grows,  
the profiles of $g_{\rm num}(x)$
approach the theoretical prediction. 
}\label{fig:gnum}
\end{center}
\end{minipage}
\hspace{2mm}
 \begin{minipage}[t]{0.48\hsize}
\begin{center}
\includegraphics[height=50mm,width=60mm]{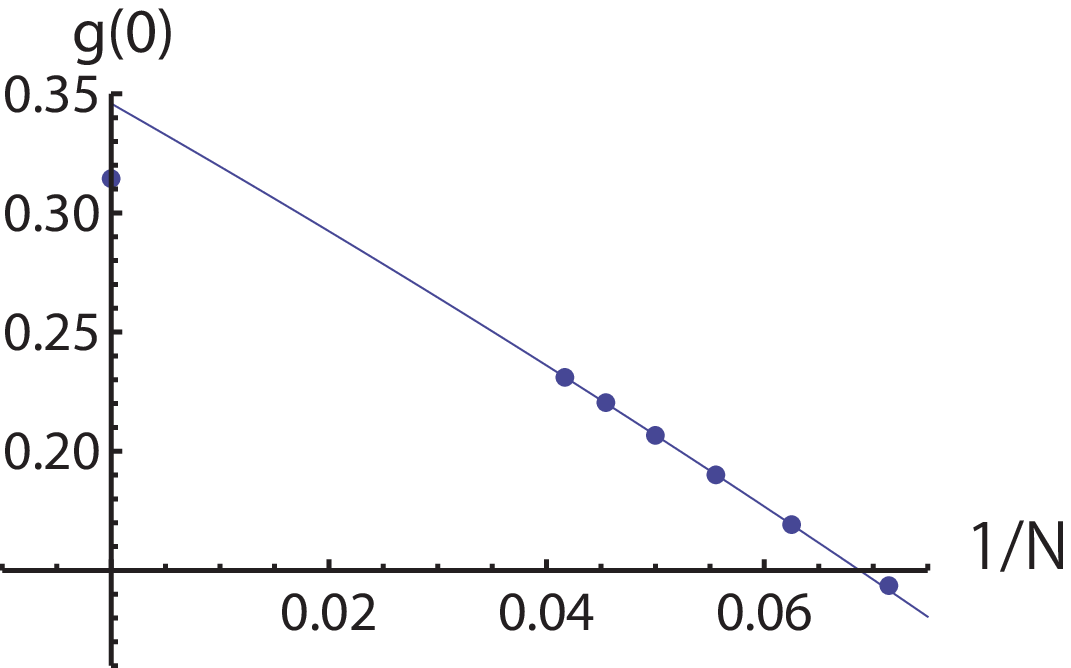}
 \caption{
Size dependence of $g_{\rm num}(0)$. The data are the same 
as figure \ref{fig:gnum}.
The solid line is provided by quadratic fitting to the data. 
The point at $1/N=0$ is the theoretical value 
derived by the replica method.
The data tend to reach this theoretical value as the system size 
grows.  
}\label{fig:g0}
\end{center}
\end{minipage}
\end{tabular}
\end{figure}

The resultant plots of $g_{\rm num}(x)$ for $\alpha=0.5$ are 
shown in figure \ref{fig:gnum}.
As the system size grows, 
the numerical data for $x\le 1$ exhibit 
flatter slopes approaching the theoretical 
prediction $g(x)=\phi^\prime_{\rm RS1}(0)$ for $x < 1$. 
This can also be seen in figure \ref{fig:g0} 
as the systematic approaching of $g_{\rm num}(0)$ 
to the theoretical value of $g(0)=0.314 \ldots$ derived from 
the replica analysis. The difference between the numerical extrapolation 
and the analytical result at $1/N=0$ is considered to be the systematic error
due to 
higher order contributions of $1/N$. 
The profiles of $x>1$, on the other hand, are approximately  
straight lines, and the slopes appear gentle
than that of the theoretical prediction $\phi_{\rm RS1}^\prime(0)$.  
However, the data still slowly move closer to 
$x \phi_{\rm RS1}^\prime(0)$ ($x>1$) as $N$ becomes larger as a whole, 
implying consistency with the theoretical prediction.

\begin{figure}[htbp]
\begin{center}
   \includegraphics[height=50mm,width=60mm]{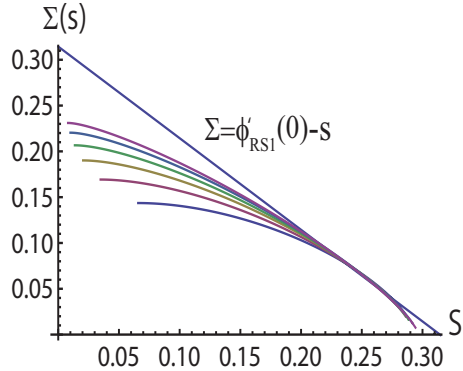}
 \caption{$\Sigma(s(x))$ obtained from $g_{\rm num}(x)$ 
using the relation of (\ref{phi_complexity}) for 
$\alpha=0.5$. The system sizes increase 
from bottom to top. The solid line denotes the asymptotic 
line in the thermodynamic limit predicted by the replica analysis.}  
\label{fig:sSigmaB}
\end{center}
\end{figure}

Complexity $\Sigma(s)$ can also be assessed from the numerical data. 
One scheme for evaluating $\Sigma(s)$ is to use the 
relations of (\ref{phi_complexity}) with a polynomial 
interpolation of the numerical data. 
We determined the order of the polynomial using Akaike's information 
criteria \cite{Akai} and eventually selected a $27$th degree polynomial, 
but the obtained results were not so sensitive to details of 
the choice of the polynomial. 
The assessed profiles of $\Sigma(s)$ are plotted in figure \ref{fig:sSigmaB}.
The curves appear to approach the line predicted in the previous 
section as $N$ increases, which supports our replica analysis. 

\begin{figure}[htbp]
\begin{tabular}{cc}
\hspace{-2mm}
\begin{minipage}[t]{0.48\hsize}
\begin{center}
 \includegraphics[height=60mm,width=70mm]{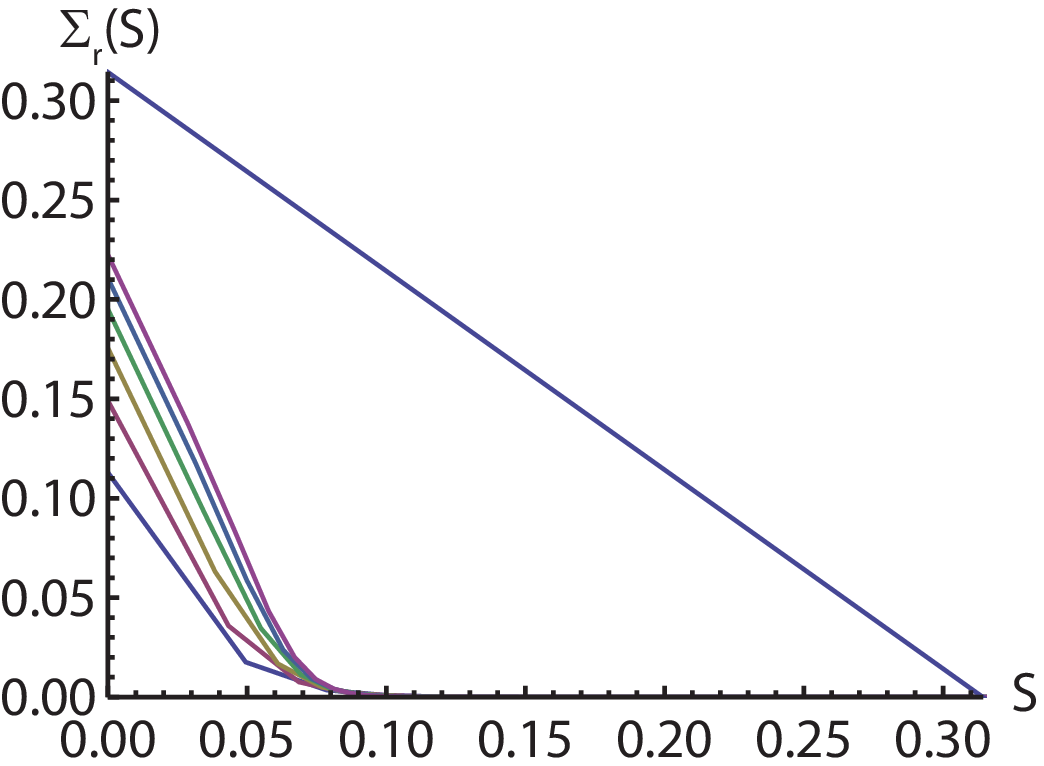}
 \caption{Plot of the raw complexity 
$\Sigma_{\rm r}(s)$ for $\alpha =0.5$. 
The system size increases from $N=14$ to $24$ in increments of 2, 
from bottom to top. The solid line is the same 
as that shown in figure \ref{fig:sSigmaB}.  
}
\label{fig:rSigma}
\end{center}
\end{minipage}
\hspace{2mm}
 \begin{minipage}[t]{0.48\hsize}
\begin{center}
\includegraphics[height=60mm,width=70mm]{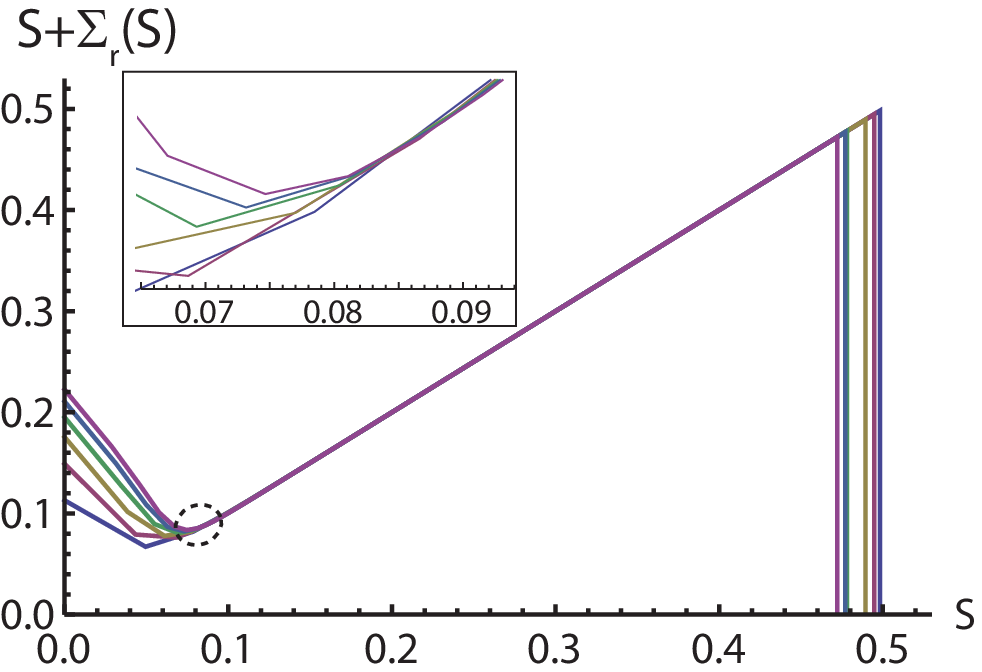}
\caption{Plot of 
$s+\Sigma_{\rm r}(s)$.  
As the system size increases, the curve appears to converge 
to a V-shape function, indicating that $\Sigma_{\rm r}(s)$ is 
convex downward. 
The inset shows a close-up of the region enclosed by the dotted ellipse. 
}
\label{fig:rspSigma}
\end{center}
\end{minipage}
\end{tabular}
\end{figure}

However, the complexity curve shown in figure \ref{fig:sSigmaB} might 
lose the information about the correct distribution of the clusters, 
as mentioned in section \ref{sec:typ}.
In order to examine this possibility, we directly evaluate 
the distribution of pure states in a rather naive manner. 
We refer to the result of this assessment as the 
{\em raw complexity}, 
which is defined as 
\begin{equation}
\Sigma_{\rm r}(s=(1/N)\log Q|D^{M})= 
\frac{1}{N}
\Theta
\left(
C(Q|D^{M})
\right)
\log
\left(
C(Q|D^{M})
\right).
\end{equation}
Taking the sample average yields the typical profile
of $\Sigma_{\rm r}( s|D^{M})$ as 
$\Sigma_{\rm r}(s)=
[\Sigma_{\rm r}( s|D^{M})]$, 
the result of which for $\alpha=0.5$ is shown in 
figure \ref{fig:rSigma}. 
We took $32,000$ samples 
in the evaluation for each size and joined the plots to obtain 
smooth curves.
This figure indicates that 
$\Sigma_{\rm r}(0)$ approaches the value of the theoretical 
prediction $\phi_{\rm RS1}^\prime(0)|_{\alpha=0.5}=0.314\ldots$ 
from below as $N$ increases. 
However, $\Sigma_{\rm r}(x)$ for $x \ge 0.1$ appears to 
remain approximately constant at zero, indicating that $\Sigma_{\rm r}(x)$ converges to a convex 
downward function.
We also plot the function $s+\Sigma_{\rm r}(s)$ in figure 
\ref{fig:rspSigma}. 
This plot shows two peaks and one dip of $s+\Sigma_{\rm r} (s)$,
indicating that $\Sigma_{\rm r} (s)$ is a convex downward function. 
The position of the right-hand peak tends to move left 
to the right terminal point $s=0.314 \ldots$
of the theoretical prediction as the system size increases, while 
the dip appears to be bounded at the point $x=0.084$
 as shown in the inset. 
In conclusion, these figures indicate
that the exponent that characterizes the size distribution 
of the pure states, $\Sigma_{\rm r}(s)$, is not 
a convex upward function in this system and does not agree with 
$\Sigma(s)$, which is evaluated by the relation of (\ref{complexity_Legendre})
using $g(x)$. 

Next, we assessed the rate function for the region of 
$\alpha>\alpha_{\rm s}$.
In this region, the generation of samples that are perfectly compatible with the Ising perceptrons rarely occurs and is dominated by $s=0$. 
Therefore, we numerically evaluated the probability that 
a given set of samples $D^M$ could be separated by the Ising 
perceptron, $P_{\rm sep}$, and estimated $R(0)$ as
$R(0)=(1/N) \log P_{\rm sep}$. 

\begin{figure}[htbp]
\begin{tabular}{cc}
\hspace{-2mm}
\begin{minipage}[t]{0.48\hsize}
\begin{center}
 \includegraphics[height=50mm,width=60mm]{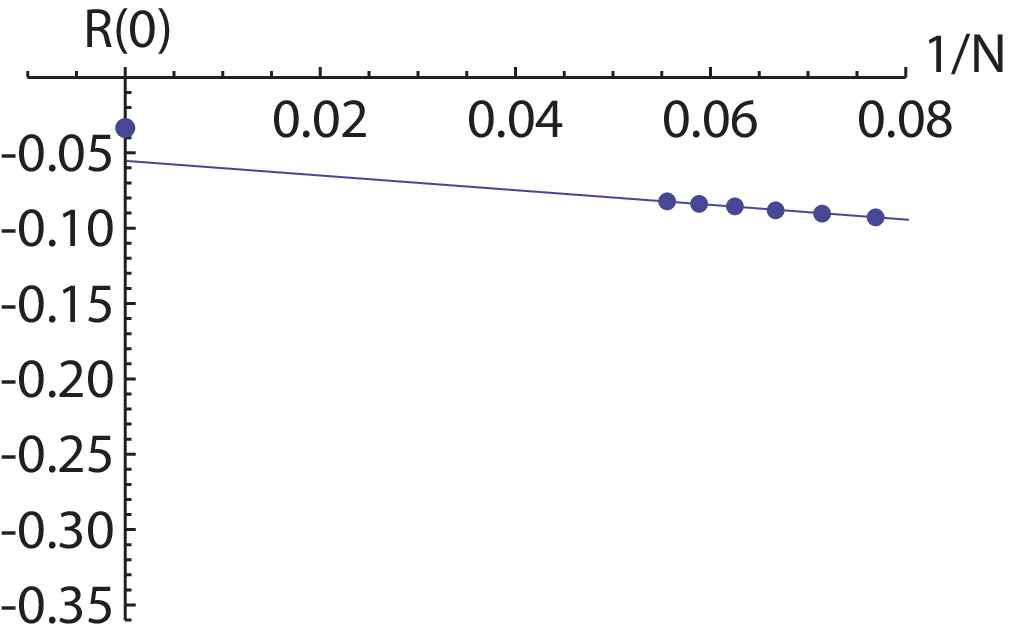}
 \caption{Size dependence of the rate function for $\alpha=1.0$. 
The point at $1/N=0$ is the value predicted by the frozen 
RSB solution. 
The system size increases from $N=12$ to $18$ in increments of $1$. 
The data from $320,000$ samples were evaluated for each $N$. 
The statistical errors are smaller than the markers.
}
\label{fig:Ra10}
\end{center}
\end{minipage}
\hspace{2mm}
 \begin{minipage}[t]{0.48\hsize}
\begin{center}
\includegraphics[height=50mm,width=60mm]{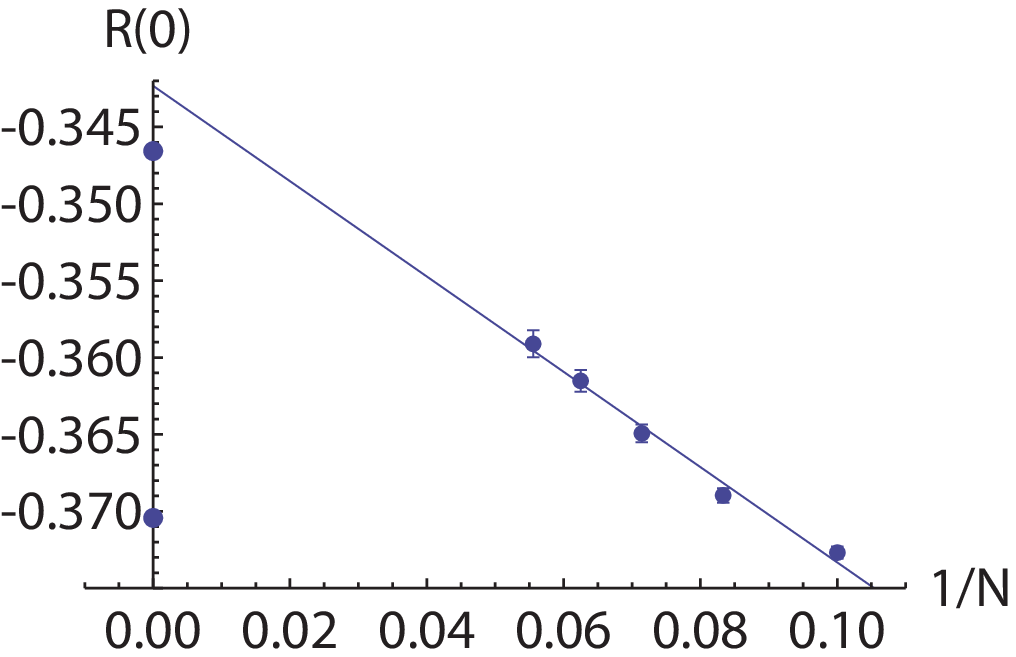}
\caption{Size dependence of the rate function for $\alpha=1.5$. 
The upper and lower plots at $1/N=0$ are given by the RS2 and 
frozen RSB solutions, respectively. 
The system size increases from $N=10$ to $18$ in increments of $2$. 
The data from $25,600,000$ samples were evaluated for each $N$. 
}
\label{fig:Ra20}
\end{center}
\end{minipage}
\end{tabular}
\end{figure}

The resultant plots are given in figures \ref{fig:Ra10} 
and \ref{fig:Ra20} for $\alpha=1.0$ and $1.5$, respectively.
The solid lines in these figures were obtained by the linear fitting 
for the numerical data. 
These figures show that the theoretical predictions
are reasonably consistent with the values of extrapolation of 
the numerical data. 
The statistical errors are sufficiently small, and hence 
the differences between the analytical and 
numerical results should be the systematic errors due to 
the nonlinearity of the Ising perceptron.

\section{Summary}
In the present paper, we investigated the structure of the weight space 
of Ising perceptrons in which a set of random patterns 
is stored using the derivatives of the generating function of 
the partition function. This was achieved by carrying out a finite-$n$ 
replica analysis under the assumption of one-step replica 
symmetry breaking (1RSB) handling Parisi's 1RSB parameter as a 
control parameter. 
For $\alpha < \alpha_{\rm s}=0.833 \ldots$, 
the analysis of $n \to 0$ indicates that 
the characteristic exponent of the size distribution of pure states 
is not convex upward, which implies that the weight space is equally 
dominated by a single large cluster of exponentially many 
weights and exponentially many clusters of a single weight. 
For $\alpha > \alpha_{\rm s}$, 
a set of random patterns is rarely compatible with the 
Ising perceptron. 
The $n\to 0$ analysis enables us to assess the rate function 
that characterizes a small probability that a cluster of 
a given entropy will emerge after the storage of random patterns.  
We found that a cluster of finite entropy is generated 
with a relatively high probability for $\alpha_{\rm s} 
< \alpha <\alpha_{\rm GD}=1.245 \ldots$, but this 
is very rare for $\alpha > \alpha_{\rm GD}$. 
These theoretical predictions have been validated by extensive 
numerical experiments. 
We also drew a complete phase diagram 
on the $n$-$\alpha$ plane, in which $(n,\alpha)=(1,\alpha_{\rm GD})$
becomes a tricritical point. The line $n=1$ that passes through 
the tricritical point is analogous to the Nishimori line in 
the theory of spin glasses. 

We stressed the use of the replica method as a tool for 
calculating the complexity and rate function.
The developed formalism enables the extraction of useful information 
about typical and atypical behaviors of the 
objective system from a single generating function in an unified manner. 
It is hoped that the results of the present study will help to clarify 
systems with complex phase spaces as well as the replica method itself.

\ack
The present study was supported in part by a Grant-in-Aid for Scientific Research on
the Priority Area ``Deepening and Expansion of Statistical Mechanical
Informatics'' from the Ministry of Education, Culture, Sports, Science
and Technology. 
One of the authors (TO) is grateful for the financial support provided through the Japan Society for the Promotion of Science (JSPS) Research Fellowship for Young Scientists program.
A portion of the numerical calculations were performed on the TSUBAME Grid Cluster at the Global Scientific Information and Computing Center (GSIC), Tokyo Institute of Technology.

\appendix
\section{Derivation of RS and 1RSB solutions}\label{sec:Deri}
For $n=1,2,\ldots, \in \mN$, the
$n$th moment of $Z$ is expressed as
\begin{eqnarray}
&&\left[Z^n \right]_{D^M}=
\sum_{\V{S}^1,\V{S}^2,\ldots,\V{S}^n} 
\left[
\prod_{a=1}^{n}\prod_{\mu=1}^{M}
\Theta 
\left(
-y_{\mu}\frac{ \V{S}^{a}\cdot\V{x}_{\mu}}{\sqrt{N}}
\right)
\right]_{D^M},
\end{eqnarray}
where the brackets $[\cdots]_{D^M}$ denote the average over the 
quenched randomness $D^M$.
The variable 
$u^{a}_{\mu}=-y_{\mu}\V{S}^{a}\cdot\V{x}_{\mu}/\sqrt{N}$ 
$(a=1,2,\ldots,n;\mu=1,2,\ldots,M)$ can 
be regarded as multivariate Gaussian random variable, 
which is characterized as 
\begin{eqnarray}
\left[u^{a}_{\mu} \right]_{D^M}=0,\,\,
\left[u^{a}_{\mu}u^{b}_{\nu} \right]_{D^M}=
\delta_{\mu \nu} 
\left (\delta_{ab}+(1-\delta_{ab})q^{ab} \right) ,\label{eq:moments^u}
\end{eqnarray}
where $q^{ab}=(1/N) \sum_{i=1}S_i^a S_i^b$ ($a,b=1,2,\ldots,n$). 
This observation yields the following expression 
\begin{eqnarray}
&&\hspace{0cm}\left[Z^n \right]_{D^M}
=\sum_{\V{S}^1,\V{S}^2,\ldots,\V{S}^n} 
\int\prod_{a<b}dq^{a b}
\delta\left(
\V{S}^{a}\V{S}^{b}-Nq^{a b}
\right)
\left[
\prod_{a=1}^{n}\prod_{\mu=1}^{M}
\Theta 
\left(
u^{a}_{\mu}
\right)
\right]_{D^M} \nonumber \\
&&\hspace{0cm}=
\int\prod_{a<b}\frac{dq^{a b}d\Wh{q}^{a b}}{2\pi}
\exp N 
\Biggl(
-\sum_{a<b} q^{a b} \Wh{q}^{a b}
+\log \sum_{S^1,S^2,\ldots,S^n}  e^{ \sum_{ a<b }\Wh{q}^{a b}S^{a}S^{b} } 
\nonumber \\
&&\hspace{7cm}+\alpha 
\log
\left[
\prod_{a=1}^{n}
\Theta 
\left(
u^a
\right)
\right]_{\V{u}}
\Biggr),\label{eq:phi} 
\end{eqnarray}
where $\left [ \cdots \right ]_{\V{u}}$ denotes the average with 
respect to the multivariate Gaussian variables the moments of which are given by 
(\ref{eq:moments^u}).
In order to derive the previous expression, 
we used the Fourier expression of the delta function
\begin{eqnarray}
\delta\left(
\V{S}^{a}\V{S}^{b}-Nq^{a b} \right )
=\int_{-{\rm i}\infty}^{+{\rm i}\infty}
\frac{d \Wh{q}^{ab}}{2 \pi}
\exp \left (\Wh{q}^{ab} (\V{S}^{a}\V{S}^{b}-Nq^{a b}) \right ). 
\end{eqnarray}
Applying the saddle-point method to (\ref{eq:phi}), 
we immediately obtain (\ref{replica_int}).
In order to investigate $\phi(n)$ for $n \in \mR$, we need an ansatz on the 
form of the saddle point $q^{ab}$. We first adopt the RS ansatz 
\begin{equation}
\Wh{q}^{ab}=\Wh{q},\,q^{ab}=q \ (a<b =1,2,\ldots,n).
\end{equation}
Under this assumption, we obtain
\begin{eqnarray}
&&\sum_{a<b} q^{a b} \Wh{q}^{a b}=\frac{1}{2}n(n-1)\Wh{q}q,\\ 
&&\sum_{S^1,S^2,\ldots,S^n}  e^{ \sum_{ a<b }\Wh{q}^{a b}S^{a}S^{b} }
=e^{-\frac{1}{2}n\Wh{q}}\int Dz (2\cosh\sqrt{\Wh{q}}z)^n.
\end{eqnarray}
Under the RS ansatz, the Gaussian variable $u^{a}$ can be decomposed to
two independent Gaussian variables of zero mean and unit variance 
$x^{a}$ and $z$ as
\begin{equation}
u^{a}=\sqrt{1-q}x^{a}+\sqrt{q}z.
\end{equation}
Using this expression, we obtain
\begin{eqnarray}
&&\left[
\prod_{a=1}^{n}
\Theta 
\left(
u^a
\right)
\right]_{\V{u}}
=
\int Dz 
\left(
E
\left(
\sqrt{\frac{q}{1-q}}z
\right)
\right)^n,
\end{eqnarray}
where $E(u)=\int_{u}^{+\infty}Dz$.
Using the above expressions, we obtain 
\begin{eqnarray}
&&\phi_{{\rm RS}}(n)=
\Extr{q,\Wh{q}}
\Biggl\{
-\frac{1}{2}n(n-1)\Wh{q}q-\frac{1}{2}n\Wh{q}+\log\int Dz (2\cosh\sqrt{q}z)^n
\nonumber
\\
&&+\alpha \log \int Dz 
\left(E
\left(
\sqrt{\frac{q}{1-q}}z
\right)
\right)^n
\Biggr\},\label{eq:phiRS}
\end{eqnarray}
The saddle point conditions are
\begin{eqnarray}
&&q=\frac{\int Dz \cosh^n \sqrt{\Wh{q}}z \tanh^2 \sqrt{\Wh{q}}z }{
\cosh^n \sqrt{\Wh{q}}z
},\label{eq:q}  \\
&&
\Wh{q}=
\frac{\alpha}{1-q}
\frac{
\int Dz \left( \frac{E'}{E}\right)^2
E^n
}
{
\int Dz
 E^{n}
},\label{eq:qhat}
\end{eqnarray}
where $E^\prime(x)=dE(x)/dx=-e^{-x^2/2}/\sqrt{2\pi}$. Note that 
the arguments of $E$ and $E'$ are $\sqrt{q/(1-q)}z$.  
As previously noted, 
there are two solutions to (\ref{eq:q}) and (\ref{eq:qhat}), 
i.e., the {\bf RS1} and {\bf RS2} solutions presented in section \ref{sec:RSsol}. 

Next, we use the 1RSB ansatz. The replica indices are divided into
$n/m$ groups of identical size $m$, and 
$q^{ab}$ and $\Wh{q}^{ab}$ are parameterized as 
\begin{equation}
(q^{ab},\Wh{q}^{ab})=
\left\{
\begin{array}{ll}
(q_{1},\Wh{q}_{1}) & (\,\, {\rm {\it a}\,\, and\,\, {\it b}\,\, belong\,\, to\,\,  the\,\, same\,\, group} \,\,)\\
(q_{0},\Wh{q}_{0}) & (\,\, {\rm otherwise}\,\,)
\end{array}
\right.
\end{equation}
This assumption yields 
\begin{eqnarray}
&&\sum_{a<b} q^{a b} \Wh{q}^{a b}=\frac{1}{2}n(m-1)\Wh{q}_{1}q_{1}+
\frac{1}{2}n(n-m)\Wh{q}_{0}q_{0}
,\\ 
&&\sum_{S^1,S^2,\ldots,S^n}  e^{ \sum_{ a<b }\Wh{q}^{a b}S^{a}S^{b} }
=e^{-\frac{1}{2}n\Wh{q}_{1}}
\int Dz_{0} 
\left(
\int Dz_{1}
(2\cosh h)^m
\right)^{n/m}
,
\end{eqnarray}
where $h=\sqrt{\Wh{q}_{1}-\Wh{q}_{0} }z_{1}+\sqrt{\Wh{q}_{0} }z_{0}$. 
The Gaussian variable $u^{a}$ can be decomposed to obtain
\begin{equation}
u^{a}=\sqrt{1-q_{1}}y_{\sigma a}
+\sqrt{q_{1}-q_{0}}x_{\sigma }+\sqrt{q_{0}}z, 
\end{equation}
where $x_{\sigma},y_{\sigma a }$ and $z$ are 
independent Gaussian variables of zero mean and unit variance.
The index $\sigma $ indicates a block 
and a pair of $\sigma$ and $a$  specifies a replica in the $\sigma$ block. 
This transformation yields the following expression:
\begin{eqnarray}
&&\left[
\prod_{a=1}^{n}
\Theta 
\left(
u^{a}
\right)
\right]_{\V{u}}
=
\int Dz 
\left(
\int Dx 
\left\{
E(y_{0}(z,x))
\right\}^m  
\right)^{n/m},
\end{eqnarray}
where 
\begin{equation}
y_{0}(z,x)=-\sqrt{\frac{q_{0}}{1-q_{1}}}z-\sqrt{\frac{q_{1}-q_{0}}{1-q_{1}}}x.
\end{equation}
Finally, we obtain 
\begin{eqnarray}
&&\phi_{{\rm 1RSB}}(n,m)
=
\Extr{q_{1},q_{0},\Wh{q}_{1},\Wh{q}_{0}}
\Biggl\{
-\frac{1}{2}n(m-1)\Wh{q}_{1}q_{1}-\frac{1}{2}n(n-m)\Wh{q}_{0}q_{0} \nonumber \\
&&
-\frac{1}{2}n\Wh{q}_{1}+
\log 
\int Dz_{0} 
\left(
\int Dz_{1} (2\cosh h)^m
\right)^{n/m} \nonumber \\
&&+\alpha \log
\int Dz_{0} 
\left(
\int Dz_{1} 
\left\{
E(y_{0}(z_{0},z_{1}))
\right\}^m  
\right)^{n/m}
\Biggr\}
.\label{eq:phi1RSB}
\end{eqnarray}
Taking the extremization, we can derive the saddle-point equations. 
The possible solutions of the equations are discussed 
in section \ref{sec:1RSBsol}.

\end{document}